\documentclass[aps,pre,reprint,groupedaddress,twocolumn]{revtex4-2}

\usepackage{graphicx}
\usepackage{dcolumn}
\usepackage{bm}
\usepackage{amsmath,amssymb,subfigure}
\usepackage{color}
\usepackage[normalem]{ulem}
\usepackage{epstopdf}
\usepackage{esint}

\newcommand{\eq}[1]{Eq.~\eqref{#1}}
\newcommand{\etal}{\textit{et al.}}
\newcommand{\reffig}[1]{Fig.~\ref{#1}}
\newcommand{\avg}[1]{\left\langle #1 \right\rangle}

\newcommand{\sr}{\mathbf{s}\cdot\mathbf{r}}
\newcommand{\sawx}{s_{x}x}
\newcommand{\szz}{s_z z}

\newcommand{\arf}{acoustic radiation force~}
\newcommand{\arp}{acoustic radiation pressure~}
\newcommand{\arm}{acoustic radiation momentum~}

\begin{document}


\title{Acoustic radiation force on small spheres due to transient acoustic fields}


\author{Antoine Riaud}
\email{antoine\_riaud@fudan.edu.cn}
\homepage[]{http://homepage.fudan.edu.cn/ariaud/}
\author{Qing Wang}
\affiliation{State Key Laboratory of ASIC and System, School of Microelectronics, Fudan University, Shanghai 200433, China}
\author{Zhixiong Gong}
\affiliation{Univ. Lille, CNRS, Centrale Lille, Univ. Polytechnique Hauts-de-France, UMR 8520 - IEMN, F-59000 Lille, France}
\author{Michael Baudoin}
\email{michael.baudoin@univ-lille.fr}
\affiliation{Univ. Lille, CNRS, Centrale Lille, Univ. Polytechnique Hauts-de-France, UMR 8520 - IEMN, F-59000 Lille, France}
\author{Jia Zhou}
\affiliation{State Key Laboratory of ASIC and System, School of Microelectronics, Fudan University, Shanghai 200433, China}


\date{\today}

\begin{abstract}
	Acoustic radiation force is a net force experienced by an object under the action of an acoustic wave. Most theoretical models require the acoustic wave to be periodic, if not purely monofrequency, and are therefore irrelevant for the study of \arf due to acoustic pulses. Here, we introduce the concept of finite-duration pulses, which is the most general condition to derive the acoustic radiation force. In the case of small spheres, we extend the Gor'kov to formula to unsteady acoustic fields such as traveling pulses and interfering wave packets. In the latter case, our study suggests that the concept of acoustic contrast is also relevant to express the acoustic radiation force. For negative acoustic contrast particles, the acoustic trapping region narrows with shorter pulses, whereas positive contrast particles (such as biological cells) can fall in secondary traps when the pulse width deviates from an optimal value. This theoretical insight may help to improve the selectivity of pulsed acoustic tweezers.  
\end{abstract}


\maketitle


\section{Introduction}

Acoustic radiation pressure is generally described as a steady force acting on surfaces exposed to acoustic waves. While radiation pressure has been extensively studied for almost two centuries, the \textit{steady} nature of the force has been little discussed. Since the pioneering work of Rayleigh \cite{rayleigh1902pressure,rayleigh1905momentum}, radiation pressure has mainly been studied in the monofrequency regime and been defined as an exchange of momentum between a wave and a particle. Integrating the \arp over an object boundaries yields a net acoustic radiation force, which has been computed for incompressible spheres exposed to plane waves by King \cite{king1934acoustic} and later on by taking the sphere compressibility into account by Yosioka and Kawasima \cite{yosioka1955acoustic}. In 1962, Gor'kov derived an elegant formula for the \arf of spheres much smaller than the acoustic wavelength subject to arbitrary acoustic fields \cite{gor1961forces}. Even though these calculations have been extended to arbitrary large spheres in complex acoustic fields \cite{baresch2013three,silva2019particle,sapozhnikov2013radiation}, and to include viscous effects and even acousto-thermal effects \cite{doinikov1997acousticI,doinikov1997acousticII,doinikov1997acousticIII,settnes2012forces,karlsen2015forces}, the monofrequency dogma has remained essentially unchallenged. 

Silva \etal~ have investigated the parametric oscillations of spheres of arbitrary size exposed to bichromatic and polychromatic acoustic waves in an inviscid fluid \cite{silva2005dynamic,silva2007multifrequency}. They have shown that, unlike the monochromatic case, the amplitude of the parametric forcing depends on the nonlinearities in the fluid equation of state. A simplified expression of the \arf was independently provided by Karlsen and Bruus for the average force experienced by small spheres exposed to polychromatic waves in a thermoviscous fluid \cite{karlsen2015forces}. In both cases, the polychromatic assumption refers to  a discrete combination of modes, which is tantamount to periodic excitation signals. Although this allows the analysis of a combination of transducers \cite{ding2012chip,tian2019wave}, the periodic assumption makes it \textit{a priori} irrelevant for continuous combinations of frequencies \cite{kang2020acoustic} and acoustic pulses \cite{collins2016acoustic}.

While the traditional use of \arf for acoustic levitation and acoustic tweezing allows considering extremely long actuation signals, and is therefore well described by existing theories, using a continuous range of frequencies offers a superior flexibility in device operation. This enables finely tuning the spatial geometry of an acoustic field, which is tantamount to adjusting the acoustic radiation force landscape. For instance, Kang \etal~have demonstrated such a multifrequency device \cite{kang2020acoustic} for the fine positioning of particles. However, one of the most interesting prospect of pulses is to increase the acoustic trapping selectivity, that is the ability to trap a particle within many others by confining the acoustic field around this particle. Unlike the complex transducers \cite{riaud2017selective,baudoin2019folding} or transducer arrays \cite{courtney2013dexterous,baresch2016observation,riaud2015anisotropic} typically needed to achieve a high selectivity, Collins \etal~have pioneered an alternative strategy using ultra-short acoustic pulses \cite{collins2016acoustic}. They assumed that, similarly to the monofrequency case, traveling acoustic pulses would generate much less acoustic force than standing acoustic pulses (obtained by the interference between two counter-propagating pulses). They demonstrated that acoustic pulse width as short as 10 acoustic periods yield an acoustic radiation force, and that acoustic interference and trapping regions overlap. Even though the limited bandwidth of their transducers restricted the pulse shortness to at least 10 periods, it not only challenges the monofrequency assumption, but also suggests that further reducing the pulse width may enable even higher selectivity.

Furthermore, non-conventional acoustic generation methods such as percussions and photoacoustic effects \cite{zharov2005photoacoustic,pezeril2011direct} can only generate brief impulsions with a broad frequency content. Starting from Longhorn's study of the effect of shock waves on small particles \cite{longhorn1952unsteady}, the hydrodynamic community has also considered the forces generated by fast variations of pressure and velocity \cite{parmar2011generalized,annamalai2017faxen}. Yet, even the most recent refinement have overlooked the effect of the sphere compressibility and describe complex dynamics on timescales that cannot be resolved experimentally.

\begin{figure}
	\includegraphics{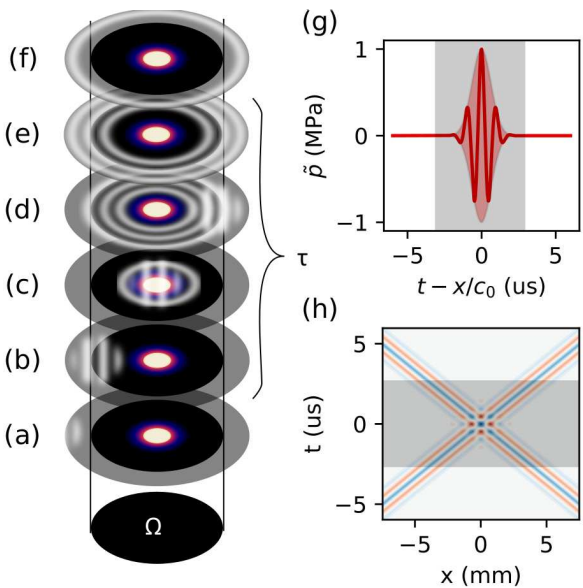}
	\caption{Scattering and interference of wave packets. \textbf{(a)-(f)} time-lapse of the diffraction of an acoustic wave by a wave packet. A particle (brown disk) surrounded by viscous and thermal boundary layers (blue and red) sits in a quiescent fluid. The fixed domain $\Omega$ is much larger than the acoustic wavelength and encompasses both boundary layers. At \textbf{(a-b)}, a wave packet enters $\Omega$, then \textbf{(c)} is scattered by the particle and \textbf{(c,d,e)} generates a scattered field (shown as circular waves) that \textbf{(f)} eventually exits $\Omega$. Neglecting the particle motion, the state of $\Omega$ is identical in steps \textbf{(a)} and \textbf{(f)} which allows defining $\tau$ as the duration between these two steps. \textbf{(g)} A Gaussian acoustic pulse. The gray region indicates the pulse duration, which can be chosen arbitrarily larger than the pulse width. \textbf{(h)} Interference of two Gaussian pulses. The gray band indicates the pulse duration chosen in this particular example.}
	\label{fig: schematic}
\end{figure}

In this paper, we question the \textit{steady} nature of the \arf by considering acoustic wave-packets of finite duration $\tau$. By finite duration, we require all the wave quantities $\tilde{x}$ to satisfy the condition $\tilde{x}(-\tau/2,\mathbf{r})$ = $\tilde{x}(\tau/2,\mathbf{r})$, where the origin of time is freely chosen. Such a wave packet is shown in \reffig{fig: schematic}.b where $\tilde{x}(-\tau/2,\mathbf{r})$ = $\tilde{x}(\tau/2,\mathbf{r})=0$. Note that the duration $\tau$ can be arbitrarily larger than the pulse width $\Theta$ (the time-window containing most of the acoustic energy of the pulse). 

This finite-duration condition is a considerably more relaxed requirement than periodic wave packets (where the equality $\tilde{x}(t-\tau/2,\mathbf{r})$ = $\tilde{x}(t+\tau/2,\mathbf{r})$ must hold \textit{for all times} $t$), and even more so for monofrequency waves. We also note that such packets allow constructing periodic wave packets and monofrequency waves by juxtaposing several periods. While the conditions on the acoustic wave are relaxed, we now need to assume that the acoustic field experienced by the particle does not change much during the entire pulse duration. Since acoustic quantities vary over a wavelength, this condition is tantamount to assuming that the particle displacement  $\Delta \ell$ is negligible compared to the shortest acoustic wavelength $\lambda$ of the excitation signal (corresponding to the highest frequency $f_{max}$ in the pulse), not only during an acoustic period, but during the entire acoustic pulse duration, that is $\frac{\Delta \ell}{\lambda}\ll l$. Similarly to previous theoretical studies on the acoustic radiation force, our calculations rely on the perturbation expansion to evaluate the force, which in turn restricts the particle migration speed $v_p\simeq\frac{\Delta \ell}{\Theta} \gtrsim \frac{\Delta \ell}{\tau} $ to $\frac{v_p}{c_0}\lesssim\epsilon^2$ \cite{karlsen2015forces}, with $\epsilon = \frac{p_{max}}{\rho_0c_0^2}$ the Mach number, where $\rho_0$, $c_0$ and $p_{max}$ the fluid density at rest, the wave speed and the peak wave pressure amplitude, respectively. This yields a coarse overestimate of this small-displacement condition:
\begin{equation}
\frac{\Delta \ell}{\lambda} \simeq \Theta f_{max}\epsilon^2 \leq \tau f_{max}\epsilon^2 \ll 1. \label{eq: max pulse width}
\end{equation}
The small $\epsilon\lesssim 0.01$ for most acoustofluidic applications suggests that the above inequalities hold as long as $\tau \lesssim 100$ acoustic periods long. Assuming that the excitation signal bandwidth is bounded by $[f_{min},f_{max}]$, this finite-pulse theory will be relevant for $f_{min}\gg f_{max}\epsilon^2 \sim 10^{-4} f_{max}$. More stringent conditions are foreseen when the particle can resonate with the incident field and emit the scattered field for a longer duration. This may occur for bubbles or particle with a size comparable to the acoustic wavelength. For periodic series of pulses, this has little consequences because the acoustic field variables will cycle over the pulse repetition period so that the \arf of each pulse period can be evaluated by carefully including the steady-state scattered field due to this periodic excitation. For isolated or stochastic acoustic pulses however, the finite duration condition would require $\tau$ large enough to allow the scattered field to decay to negligible levels before evaluating the acoustic radiation force.

This finite duration allows to overlook the complex dynamics found by Longhorn and subsequent studies \cite{longhorn1952unsteady,parmar2011generalized,annamalai2017faxen,silva2005dynamic,silva2006parametric,silva2007multifrequency}, which are difficult to resolve experimentally at present \cite{chen2005measurement,silva2006parametric}, and to focus instead on the transformation of the particle from the initial state before wave packet scattering and a final state after the scattering event. This is illustrated in \reffig{fig: schematic}.a-f: the domain $\Omega$ contains a particle (brown disc) with a thermo-viscous boundary layer (in blue and red). An incident wave packet falls on the particle and generates a scattered field that eventually escapes $\Omega$ at step (f). The finite-duration condition requires that the state of $\Omega$ at step (f) is the same as in step (a).

After generalizing the expression of the acoustic radiation pressure tensor to finite duration wave packets, we will re-derive the \arf on small spheres while relaxing the monofrequency assumption. This will provide a generalized Gor'kov equation for acoustic wave packets of finite duration. Although our final expression could be derived from Karlsen \etal~expression for periodic waves \cite{karlsen2015forces} using Parseval's theorem, and even for non-periodic pulses using Plancherel theorem, (i) re-deriving the equation provides a deeper physical meaning of its origin, especially regarding the scattering coefficients, and (ii) our derivation has a broader set of validity than what these theorems alone would warrant. In particular, using the Plancherel theorem requires to integrate the time-domain function over $[-\infty,+\infty]$ and would therefore not ensure, for instance, that successive pulses would not interfere. After deriving this general expression, we will then evaluate the \arf of these packets in the simplified cases of traveling and standing waves, and use our analytical expression to optimize the selectivity of acoustic trapping by acoustic pulses.

 \section{Model}
 
\subsection{Definition of the \arf}
Using the summation of repeated indices, mass and momentum conservation in a Newtonian fluid of shear viscosity $\mu$ and bulk viscosity $\mu_b$ read:
\begin{subequations}
	\begin{align}
	\partial_t \rho + \partial_i \rho v_i &= 0, \label{eq: mass cons} \\
	\partial_t \rho v_i + \partial_j \rho v_i v_j &= \partial_j \left(-p\delta_{ij} +  \Sigma_{ij}\right) \label{eq: momentum cons}\\ 
	\Sigma_{ij} &= \mu\left(\partial_{j}v_i +\partial_{i}v_j \right)+\left(\frac{\mu}{3}+\mu_b\right)\delta_{ij}\partial_{j}v_j, \label{eq: stress}
	\end{align}
\end{subequations}
with $p$, $\rho$ and $\mathbf{v}$ the fluid pressure, density and velocity, and $\delta_{ij}$ is the Kronecker delta function. These equations are complemented with the isentropic equation of state up to the second order in $\rho-\rho_0$:
\begin{equation}
p - p_0 = {c_0}^2(\rho-\rho_0)+ \frac{1}{2}\Gamma(\rho-\rho_0), \label{eq: eos}
\end{equation}
with $\Gamma = \frac{B}{A}\frac{{c_0}^2}{\rho_0}$, where $\frac{B}{A}$ is an important parameter in nonlinear acoustics \cite{silva2005dynamic,baudoin2020acoustic}.

A particle with boundaries located at $\partial\Omega_p(t)$ immersed in this fluid will experience a force $F_i = \oint_{\partial\Omega_p(t)}p\delta_{ij}-\Sigma_{ij}dS_j$ \cite{karlsen2015forces,baudoin2020acoustic}. When the fluid motion is solely due to acoustic waves, the time-average of this force is the acoustic radiation force. However, the particle position and shape are affected by the acoustic field, meaning that the integration boundary $\partial\Omega_p$ varies with time. This problem is addressed by using the Gauss integral theorem on \eq{eq: momentum cons} to get the Reynolds transport equation. This allows to evaluate the stress from a fixed boundary $\partial_\Omega$ located arbitrarily far away from the particle:
\begin{equation}
\partial_t \int_\Omega \rho v_i dV + \oint_{\partial\Omega} \left( \rho v_i v_j +p\delta_{ij} -\Sigma_{ij} \right)dS_j =  -\mathcal{F}^{tot}_i \label{eq: Reynolds}
\end{equation}
with $\mathcal{F}^{tot}_i$ the reaction force of the fluid to all the external forces acting on it. 

The main caveat of using the transport theorem is that it operates a momentum balance over the entire domain $\Omega$, such that acoustic momentum transfer from the wave to the fluid inside $\Omega$, and the steady flow generated by the acoustic wave outside this domain (acoustic streaming \cite{riaud2017influence}) are included in the left-hand term \cite{baudoin2020acoustic}. It is therefore customary to distinguish the variations of momentum due to the particle itself (acoustic scattering, microstreaming and so on) and the effects that arise from the acoustic wave attenuation that would have occurred even in the absence of any particle. Conveniently, the spatial scale of acoustic attenuation $\Lambda_{ac} = \frac{\rho_0{c_0}^3}{\mu\left(4/3+\frac{\mu_b}{\mu}\right)\omega^2}$ is often many orders of magnitude larger than the scale of the viscous $\Lambda_{visc} = \sqrt{2\frac{\mu}{\rho_0\omega}}$, and thermal  $\Lambda_{therm} = \sqrt{2\frac{D_T}{\omega}}$ boundary layers (with $D_T$ the thermal diffusivity in the fluid), which suggests that the radiation force is well approximated by considering an inviscid and adiabatic fluid, provided that the local effects of microstreaming and heat conduction are well resolved in the direct vicinity of the particle \cite{karlsen2015forces}. Using this inviscid and adiabatic approximation to compute the acoustic radiation force, the viscous effects can be recovered by computing the acoustic streaming and adding it as a drag force acting on the particle \cite{muller2013ultrasound}.

\subsection{Perturbation expansion in the far field}

Based on the discussion above, we now omit the viscous effects in the fluid, and use the small Mach number $\epsilon$ to expand the evolution of the pressure, velocity and density fields by perturbation of increasing orders in $\epsilon$. 0-order quantities are denoted $x_0$, first order $\tilde{x}$ and second-order $\bar{x}$:
\begin{subequations}
	\begin{align}
	\rho &= \rho_0 + \epsilon\tilde{\rho}+\epsilon^2\bar{\rho},\\
	p &= p_0 + \epsilon \tilde{p}+\epsilon^2 \bar{p},\\
	v_i &= 0 + \epsilon \tilde{v}_i+\epsilon^2 \bar{v}_i.
	\end{align}
\end{subequations}

In order to get non-trivial values for the Mach number, we now solve each perturbation order independently. The order 0 in $\epsilon$ is the hydrostatic equilibrium:
\begin{subequations}
	\begin{align}
	\partial_t \rho_0 &= 0, \\
	\partial_i p_0 &= 0, 
	\end{align}
\end{subequations}
which suggests a uniform pressure $p_0$ and density $\rho_0$ in the fluid at rest. Then, the 1st order in $\epsilon$ describes the acoustic field:
\begin{subequations}
	\begin{align}
	\partial_t \tilde{\rho} + \rho_0 \partial_i \tilde{v}_i &= 0, \label{eq: mass cons1} \\
	\rho_0  \partial_t \tilde{v}_i &= -\partial_i \tilde{p}. \label{eq: momentum cons1}
	\end{align}
\end{subequations}
Taking the divergence of \eq{eq: momentum cons1} and using the equation of state (\eq{eq: eos}) up to the first order in $\epsilon$ yields the d'Alembert equation:
\begin{equation}
	\partial^2_{tt} \tilde{p} - {c_0}^2\partial^2_{ii} \tilde{p} = 0. \label{eq: wave}
\end{equation} 

The second order in $\epsilon$ unveils the main nonlinear effects:
\begin{subequations}
	\begin{align}
	\partial_t \bar{\rho} + \partial_i \tilde{\rho}\tilde{v}_i + \rho_0 \partial_i \bar{v}_i &= 0, \label{eq: mass cons2} \\
	\partial_t \rho_0 \bar{v_i} + \partial_t \tilde{\rho} \tilde{v_i} + \rho_0 \partial_j  \tilde{v}_i \tilde{v}_j &= -\partial_i \bar{p}, \label{eq: momentum cons2}\\
	\bar{p} = {c_0}^2\bar{\rho} + \frac{1}{2}\Gamma{\tilde{\rho}}^2. \label{eq: eos2}
	\end{align}
	\label{eq: order2}
\end{subequations}

To facilitate comparison with earlier works \cite{gor1961forces,settnes2012forces,karlsen2015forces}, we will consider the average of the quantities of interest, instead of the time-integral as suggested by the physical setting: $\avg{x} = \frac{1}{\tau}\int_{-\tau/2}^{\tau/2}x(t)dt$. Due to the finite duration of the pulse, any 1st order quantity $\tilde{x}$ satisfies:
\begin{equation}
\avg{\partial_t \tilde{x}} = \tilde{x}(\tau/2) - \tilde{x}(-\tau/2) = 0,
\end{equation} 
This allows to simplify Eqs (\ref{eq: mass cons2}, \ref{eq: momentum cons2}, \ref{eq: eos2}):
\begin{subequations}
	\begin{align}
	\partial_i \avg{\tilde{\rho}\tilde{v}_i} + \rho_0 \partial_i \avg{\bar{v}_i} &= 0, \label{eq: avg mass cons2} \\
	\rho_0 \avg{\partial_j  \tilde{v}_i \tilde{v}_j} &= -\partial_i \avg{\bar{p}}, \label{eq: avg momentum cons2}\\
	\bar{p} &= {c_0}^2\bar{\rho} + \frac{1}{2}\Gamma\avg{\tilde{\rho}}^2. \label{eq: avg eos2}
	\end{align}
\end{subequations}
Substituting Eqs (\ref{eq: mass cons1}, \ref{eq: momentum cons1}) into $\rho_0 \avg{\partial_j  \tilde{v}_i \tilde{v}_j}$ and then using integration by part, we get:
\begin{equation}
	\rho_0 \avg{\partial_j  \tilde{v}_i \tilde{v}_j} = \partial_i \avg{\mathcal{L}}, \label{eq: L}
\end{equation}
with $\mathcal{L} = \mathcal{K}-\mathcal{V}$ the Lagrangian density of the wave, with the acoustic kinetic energy $\mathcal{K}=\frac{1}{2}\rho_0\tilde{v}_j\tilde{v}_j$ and the acoustic potential energy $\mathcal{V}=\frac{\tilde{p}^2}{2\rho_0 {c_0}^2}$.

\subsection{Lagrangian pressure and Brillouin tensor}
In order for the fluid to be at mechanical equilibrium, one must satisfy \eq{eq: avg momentum cons2}. Substituting \eq{eq: L} in \eq{eq: avg momentum cons2} yields the Lagrangian pressure:
\begin{equation}
	\avg{\bar{p}} = \mathcal{C} -\avg{\mathcal{L}}, \label{eq: lagrangian pressure}
\end{equation}
with $\mathcal{C}$ a constant independent of the position in the fluid \cite{hasegawa2000general}. Similarly to the monofrequency case \cite{hasegawa2000general,baudoin2020acoustic}, a consequence of \eq{eq: lagrangian pressure} is to set the value of the density $\bar{\rho}$ to fulfill \eq{eq: avg eos2} so that the nonlinearities in the equation of state play no role in the average acoustic radiation pressure of acoustic pulses. We note that this assertion is valid only over the whole duration of the pulse, while it was shown that $\Gamma$ was important for the detailed dynamics \cite{silva2006parametric}.

Taking the time-average of \eq{eq: Reynolds} and expanding up to the second order in $\epsilon$, we get:
\begin{equation}
\oint_{\partial\Omega} \avg{\mathcal{B}_{ij}} dS_j =  -\mathcal{F}_i, \label{eq: Reynolds avg}
\end{equation}
with $\avg{\mathcal{B}_{ij}} = \avg{\rho_0\tilde{v}_i \tilde{v}_j - \mathcal{L}\delta_{ij}}$ the Brillouin tensor, which remains the same as in the monofrequency regime. \eq{eq: Reynolds avg} is the starting point to compute the \arf on arbitrarily-shaped objects \cite{gong2019t} without additional restrictions of size other than satisfying the finite-duration conditions detailed previously.

\subsection{Acoustic radiation force on small spheres}

It is worth noting that, had \eq{eq: L} been valid everywhere inside $\Omega$, \eq{eq: Reynolds avg} would vanish. This suggests to decompose the acoustic field quantities $\tilde{x}$ into a background incident acoustic field, denoted by the subscript $\tilde{x}_{in}$, (that fulfills \eq{eq: L} and therefore generates no force); and a scattered acoustic field due to the particle, denoted by the subscript $\tilde{x}_{sc}$:

\begin{subequations}
	\begin{align}
	\tilde{\rho} &= \tilde{\rho}_{in}+ \tilde{\rho}_{sc},\\
	\tilde{p} &= \tilde{p}_{in}+ \tilde{p}_{sc},\\
	\tilde{v}_i &= \tilde{v}_{i,in}+ \tilde{v}_{i,sc},\\
	\tilde{\phi} &= \tilde{\phi}_{in}+ \tilde{\phi}_{sc}, \label{eq: phi}
	\end{align}
\end{subequations}
where we introduced the convenient potential $\tilde{\phi}$, with $\tilde{v}_i = \partial_i\tilde{\phi}$ and $\tilde{p} = -\rho_0\partial_t\tilde{\phi}$. 

In order to evaluate the scattered field, it is convenient to work in the frequency domain. Any real time-varying functions $\tilde{x}$ can be decomposed into its monofrequency components $\hat{x}=\check{x}(\omega)e^{-i\omega t}$:
\begin{equation}
\tilde{x} = \frac{1}{2}\int_{0}^{\infty}(\hat{x}+\hat{x}^\ast) d\omega. \label{eq: Fourier}
\end{equation}

So far, we have only assumed that either the particle is non-resonant or that (i) the acoustic excitation has some periodicity or (ii) the time separation between pulses is long enough to let the scattered field radiated by the particle decay to negligible levels. We now assume that the particle radius $a$ is much smaller than the acoustic wavelength, which yields relatively simple expressions for the scattered field measured at a distance $r\gg\lambda$:
\begin{equation}
\hat{\phi}_{sc} = -f_1(\omega)\frac{a^3}{3\rho_0}\frac{\partial_t \left.\hat{\rho}_{in}\right|_p}{r}
-f_2(\omega)\frac{a^3}{2}\partial_i\left[
\frac{\left.\hat{v}_{i,in}\right|_p}{r}\right], \label{eq: scattered}
\end{equation}
where $x|_p$ indicates that $x$ is evaluated at the location of the particle ($\mathbf{0}$) at the retarded time $(t-r/c_0)$. We note that these functions still depend on space due to the time-retarded argument. The monopole and dipole scattering coefficients $f_1$ and $f_2$ are complex numbers which depend on the chosen convention $\hat{x}=\check{x}(\omega)e^{-i\omega t}$. They can be found in the comprehensive study by Karlsen and Bruus \cite{karlsen2015forces}.

Substituting \eq{eq: phi} in \eqref{eq: Reynolds avg}, and neglecting squares of $\tilde{\phi}_{in}$ as they model a wave without any interaction with a particle, therefore yielding no momentum exchange, and the squares of $\tilde{\phi}_{sc}$ proportional to $a^6$ and therefore negligible for a small particle, we get:

\begin{equation}
\mathcal{F}_i = -\int_\Omega \rho_0\avg{\tilde{v}_{i,in}\left(\partial^2_{jj}-\frac{1}{{c_0}^2}\partial^2_{tt}\right)\tilde{\phi}_{sc}}dV. \label{eq: F_up_to_a6}
\end{equation}

Similarly to the previous studies \cite{gor1961forces,settnes2012forces,karlsen2015forces}, we recall that the D'Alembert equation acting on the monopole and dipole terms  \eqref{eq: wave} result in a singular density point $\left.\hat{\rho}_{in}\right|_p\delta(\mathbf{r})$ and a singular velocity point $\left.\hat{v}_{i,in}\right|_p\delta(\mathbf{r})$, with $\delta$ the Dirac distribution. 

\begin{multline}
\left(\partial^2_{jj}-\frac{1}{{c_0}^2}\partial^2_{tt}\right)\hat{\phi}_{sc} = f_1(\omega)\frac{4\pi a^3}{3\rho_0}\partial_t \left.\hat{\rho}_{in}\right|_p\delta(\mathbf{r})\\
+ f_2(\omega) 2\pi a^3 \partial_j \left.\hat{v}_{j,in}\right|_p\delta(\mathbf{r}). \label{eq: phihat_sc}
\end{multline}
Integrating \eq{eq: phihat_sc} over all angular frequencies $\omega$, we get the time-dependent scattered field:
\begin{multline}
\left(\partial^2_{jj}-\frac{1}{{c_0}^2}\partial^2_{tt}\right)\tilde{\phi}_{sc} = \\ 
\frac{4\pi a^3}{3}\left[\frac{1}{\rho_0 {c_0}^2}\partial_t \delta(\mathbf{r}) \tilde{p}_{m}|_p
+ \partial_j \delta(\mathbf{r}) \tilde{v}_{j,d}|_p\right], \label{eq: phitilde_sc}
\end{multline}
with the monopole and dipole scattered fiels:
\begin{eqnarray}
\tilde{p}_{m}|_p &=& \frac{1}{2}\int_{0}^\infty f_1(\omega) \left.\hat{p}_{in}\right|_p  + f_1^\ast(\omega) \left.\hat{p}^\ast_{in}\right|_p   d\omega,
\label{eq: pm}\\
\tilde{v}_{j,d}|_p &=& \frac{3}{4}\int_{0}^\infty f_2(\omega) \left.\hat{v}_{j,in}\right|_p  + f_2^\ast(\omega) \left.\hat{v}_{j,in}^\ast\right|_p  d\omega,
\label{eq: vd}
\end{eqnarray}
where the $\partial_t$, $\partial_j$ operators and the $\delta$ distribution commute with the integral over angular frequencies.

Substituting \eq{eq: phihat_sc} in \eq{eq: Reynolds avg} and using Gauss theorem (see \cite{settnes2012forces} for additional details), the time-retarded argument in the scattered fields is simplified by the Dirac function, so that  Eqs. (\ref{eq: pm}, \ref{eq: vd}) can now be evaluated at the location of the particle. We obtain a first expression of the acoustic radiation force:
\begin{equation}
\mathcal{F}_i =  -\frac{4\pi a^3}{3}\left[\avg{\tilde{v}_{i,in}\frac{1}{{c_0}^2}\partial_t \tilde{p}_{m}} - \avg{\rho_0\tilde{v}_{j,d}\partial_j \tilde{v}_{i,in}}\right], \label{eq: F first}
\end{equation}
Integrating \eq{eq: F first} by part then yields:
\begin{equation}
\mathbf{\mathcal{F}} =  -\frac{4\pi a^3}{3}\left[\avg{\frac{1}{\rho_0{c_0}^2}\tilde{p}_{m}\nabla \tilde{p}_{in}} - \avg{\rho_0 \tilde{\mathbf{v}}_{d} \cdot\nabla \tilde{\mathbf{v}}_{in} }\right], \label{eq: F}
\end{equation}
which in the monofrequency regime simplifies into the Gor'kov equation \cite{gor1961forces,settnes2012forces,karlsen2015forces}.

\section{Results and discussion}

In the following, we consider the effect of pulses on the motion of particles. While \eq{eq: F} is valid for any finite-duration pulse, the general calculation is complicated by the convolution products in Eqs. (\ref{eq: pm}, \ref{eq: vd}). These equations can be simplified (i) by considering narrow-bandwidth acoustic beams so the variation of all quantities but the wave spectrum can be neglected over the integration bandwidth, or (ii) by considering inviscid fluids where the scattering coefficients can be factored out of the integrals.

When a particular example is needed, we will use Gaussian plane wave packets propagating along the $x$ direction $\tilde{p}_{in}(\theta) = p_{max} w(\theta)$ with:
\begin{equation}
w(\theta) = \exp(-\sigma\theta^2)\cos(\omega \theta), \label{eq: wavelet}
\end{equation}
with $\theta = t-\sr$ the retarded time of the wave, where $\mathbf{s} = \frac{1}{c_0}\mathbf{e}_x$. Advantageously, calculations with these waves are relatively straightforward but still allow tuning the pulse width ($\Theta \propto \sigma^{-1/2}$) by changing $\sigma$. For such traveling plane waves, the velocity field is given by $\tilde{v}_{in} = \tilde{p}_{in}\frac{1}{\rho_0 c_0}\mathbf{e}_x$. An important quantity when studying the \arf is the mean energy density $\avg{\mathcal{E}}$, defined as the total energy flux per unit length $\avg{\mathcal{E}} = \mathbf{s}\cdot\avg{\tilde{p}\tilde{\mathbf{v}}}$. In the case of a Gaussian wave packet, it reads:
\begin{subequations}
	\begin{align}
	&\avg{\mathcal{E}} =\avg{\mathcal{E}_\infty} \left(1+e^{-\frac{\omega^2}{2\sigma}}\right),  \label{eq: E} \\
	\text{with\quad} & \avg{\mathcal{E}_\infty} =\frac{\tilde{p}^2_{max}}{\rho_0{c_0}^2}\frac{1}{\tau}\sqrt{\frac{\pi}{8\sigma}}.
	\label{eq: E_inf}
	\end{align}
\end{subequations}
The time and frequency spread of the wave packet (defined as the standard deviation in time and frequency space) will also be useful, and read (respectively):
\begin{subequations}
	\begin{align}
	\delta t &=\frac{1}{\sqrt{2\sigma}},  \label{eq: dt} \\
	\delta f &=\sqrt{\frac{\sigma}{2\pi^2}}.
	\label{eq: df}
	\end{align}
\end{subequations}

\subsection{Narrow-band waves in a thermoviscous fluid}

We first consider an acoustic wave packet with a bandwidth $\delta f = \frac{\delta\omega}{2\pi}$ narrow enough to neglect the variations of $f_1$ and $f_2$ over this frequency band. Before discussing the force, we note that this assumption poses some constraints on the pulse duration. According to Settnes, Karlsen and Bruus, this requires that the thermoviscous boundary layer thickness $\Lambda_{tv}$ to particle radius ratio do not change much over $\delta \omega$ \cite{settnes2012forces,karlsen2015forces}, which yields $\delta\omega \ll 2\omega\frac{a}{\Lambda_{tv}}$, with $\Lambda_{tv}=\min(\Lambda_{visc},\Lambda_{therm})$. According to the Gabor limit, $\delta t\delta f \geq \frac{1}{2\pi}$ \footnote{This result differs from the well-known Gabor-Heisenberg result $\delta t\delta f \geq \frac{1}{4\pi}$ obtained for the uncertainty on density of states as opposed to wave functions. In the former case, evaluating the density of states raises the distributions to the power 2, thereby reducing their spread by half.}, with the equality being true when using Gaussian wave packets. Combining these inequalities with \eq{eq: dt} yields a lower limit for the number of periods $n_T$ of the pulse:
\begin{equation}
n_T \simeq \sqrt{\frac{2\omega^2}{\sigma}} \gg \frac{\Lambda_{tv}}{a}.
\end{equation}
While this inequality does not contradict \eq{eq: max pulse width}, their combination restricts this thermoviscous case to low-amplitude long-duration wave packets.

Separating the real and imaginary parts of $f_1 = f_1^r+if_1^i$ and $f_2= f_2^r+if_2^i$ and noting that in the current complex convention for the scattering coefficients we have $i(\hat{x}-\hat{x}^\ast) = -\frac{1}{\omega}\partial_t(\hat{x}+\hat{x}^\ast)$, yields:
\begin{eqnarray}
\tilde{p}_{m}|_p &=& f_1^r \left.\tilde{p}_{in}\right|_p - \frac{f_1^i}{\omega} \partial_t \left.\tilde{p}_{in}\right|_p, \label{eq: p_m monofreq}\\
\tilde{v}_{j,d}|_p &=& \frac{3}{2} \left(f_2^r \left.\tilde{v}_{j,in}\right|_p - \frac{f_2^i}{\omega} \partial_t \left.\tilde{v}_{j,in}\right|_p \right), \label{eq: v_d monofreq}
\end{eqnarray} 
where, consistently with the narrow-band hypothesis, the variations of $1/\omega$ over the frequency interval were neglected.

Substituting Eqs. (\ref{eq: p_m monofreq}, \ref{eq: v_d monofreq}) in \eq{eq: F}, and omitting the real part of $f_1$ and $f_2$ that do not contribute to the \arf for traveling waves up to $O(a^6)$ \footnote{See the case of traveling waves in an inviscid fluid for more details.}, we get the force due to a narrow-band traveling wave $\tilde{p}(\theta)$ in a thermoviscous fluid: 
\begin{equation}
\mathbf{\mathcal{F}} = -\mathbf{s}\frac{4\pi a^3}{3}\frac{1}{\omega\rho_0{c_0}^2}\left[f_1^i - \frac{ 3 f_2^i}{2}\right]\avg{(\partial_\theta\tilde{p}_{in})^2},
\end{equation}
 which agrees with \cite{settnes2012forces} except for a minor typo in the sign of $f_1^i$.
 
In the case of Gaussian wave packets, the force simplifies into:
\begin{equation}
\mathbf{\mathcal{F}} =  -\mathbf{s}\omega\frac{4\pi a^3}{3}\left[f_1^i - \frac{ 3 f_2^i}{2} \right] \avg{\mathcal{E}_\infty}. \label{eq: narrowband}
\end{equation}
Careful inspection of \eq{eq: narrowband} reveals that the \arf decreases in $1/\tau$. This mathematical artifact is due to the time-averaging of the force over the time $\tau$ chosen arbitrarily. Hence, a better measure of the effect of the \arf of finite-duration pulses is the transmitted \arm $\mathbf{q} = \tau\mathbf{\mathcal{F}}$, which depends on the physically-relevant pulse width ($\Theta \propto \sigma^{-1/2}$) but is independent of the pulse duration.

\subsection{Arbitrary waves in an inviscid fluid}

When the particle is much larger than the visco-acoustic or thermo-acoustic boundary layers, the scattering coefficients $\check{f}_1(\omega) = f_1^r$ and $f_2(\omega)=f_2^r$ are purely real and independent of the frequency \cite{gor1961forces}, and therefore can be factored out of the integrals in Eqs. (\ref{eq: pm}, \ref{eq: vd}), which yields:
\begin{equation}
\mathbf{\mathcal{F}} =  -\frac{4\pi a^3}{3}\nabla \avg{\mathcal{U}} \label{eq: F_inviscid}
\end{equation}
with the dynamic Gor'kov potential:
\begin{equation}
\mathcal{U} = \frac{f_1}{2\rho_0{c_0}^2} \tilde{p}_{in}^2 - \frac{3\rho_0 f_2}{4} {\tilde{\mathbf{v}}_{in}}^2. \label{eq: U} 
\end{equation}

\subsubsection{Traveling waves}

In the case of traveling waves, $\tilde{p}_{in}$ and $\tilde{v}_{i,in}$ are only functions of $\theta = t-\sr$ with $\mathbf{s}$ the wave slowness. Therefore, $\mathcal{U}$ is a function of $\theta$ only, such that:
\begin{equation}
\nabla\avg{\mathcal{U}} = \avg{\partial_\theta\mathcal{U}\nabla \theta}= -\mathbf{s}\avg{\partial_t\mathcal{U}} = O(a^6), \label{eq: traveling}
\end{equation}
which vanishes up to the small terms in $O(a^6)$ that were neglected in \eq{eq: F_up_to_a6}. Extrapolating from the monofrequency regime, one may reasonably expect that higher order terms called scattering force will dominate the \arf \cite{baresch2016observation,sapozhnikov2013radiation} in this case.

\subsubsection{Interference of two plane wave packets}

\begin{figure}
	\includegraphics{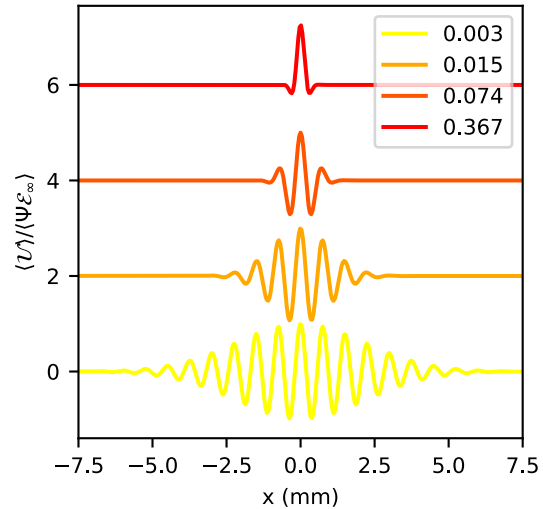} 
	\caption{Normalized Gor'kov potential $\frac{\avg{\mathcal{U}}}{\avg{\Psi\mathcal{E}_\infty}}$ from the interference of two wave packets for different values of $\sigma/\omega^2$ in water (acoustic frequency $1$ MHz, sound speed $c_0=1500$ m/s). The vertical offset between curves is added to improve clarity.}
	\label{fig: U}
\end{figure}

\begin{figure}
	\includegraphics{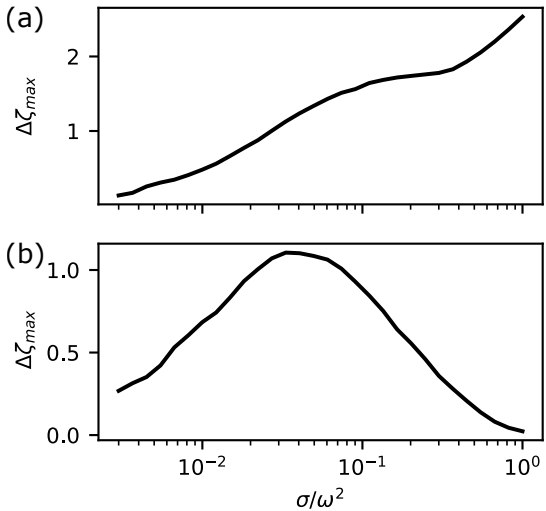} 
	\caption{Selectivity of the \arf obtained during the interference of two wavepackets of various dimensionless widths $\sigma/\omega^2$. \textbf{(a)} Negative acoustic contrast $\Psi$. \textbf{(b)} Positive acoustic contrast $\Psi$. The selectivity $\Delta\zeta_{max}$ measures the difference of restoring force between primary and secondary traps.}
	\label{fig: selectivity}
\end{figure}

We next consider two interfering plane wave packets  $\tilde{p} = \tilde{p}(\theta^+) + \tilde{p}(\theta^-)$, as illustrated in \reffig{fig: schematic}.c. For the sake of generality, the two packets intersect with an angle $2\eta_R$, so that $\theta^+ = t +\sawx-\szz$ and $\theta^- = t-\sawx-\szz$, with $s_x = \frac{\sin\eta_R}{c_0}$ and $s_z = \frac{\cos\eta_R}{c_0}$. While $\eta_R=\pi/2$ is relevant for a frontal interference \cite{mandralis1993fractionation,lenshof2012acoustofluidics}, the general case applies to surface-acoustic-wave-based tweezers such as the ones proposed by Collins \etal~ where $\eta_R$ plays the role of the Rayleigh angle that the acoustic radiation makes with the substrate \cite{vanneste2011streaming}.

Assuming that the incident wave reads $\tilde{p}_{in} = p_{max}\left[\tilde{w}(\theta^+)+\tilde{w}(\theta^-)\right]$, with $w$ given in \eq{eq: wavelet}, we have $\tilde{v}_{in,x} = -\frac{p_{max}s_x}{\rho_0 c_0}\left[\tilde{w}(\theta^+)-\tilde{w}(\theta^-)\right]$ and $\tilde{v}_{in,z} = \frac{\tilde{p}_{in}s_z}{\rho_0}$. This yields the Gor'kov potential and the \arm $\mathbf{q}$ transferred by an acoustic wave during the interference of 2 wavelets:

\begin{subequations}
	\begin{align}
	\mathcal{U} &= \avg{\mathcal{E}_\infty}\Psi\left[\cos(2\omega \sawx)+e^{-\frac{\omega^2}{2\sigma}}\right]G(\sawx),\\
	\mathbf{q} &= -2\omega s_x\frac{4\pi a^3}{3}\tau \Psi\avg{\mathcal{E}_\infty}\zeta(\sawx)\mathbf{e}_x,\\
	 \Psi &= f_1^r-\cos(2\eta_R)\frac{3}{2}f_2^r,
	\end{align}
\end{subequations}
with $G(\sawx) = e^{-2\sigma(\sawx)^2}$. The acoustic contrast factor is identical to the monofrequency case \cite{simon2017particle}. The dimensionless restoring force $\zeta$ reads:
\begin{multline}
\zeta(\sawx) = 2\left[\frac{2\sigma\sawx}{\omega}(e^{-\frac{\omega^2}{2\sigma}}+\cos(2\omega \sawx))\right.\\
+\left.\sin(2\omega \sawx)\right]G(\sawx)
\end{multline}
Examples of Gor'kov potential for various values of $\sigma/\omega^2$ are shown in \reffig{fig: U}. For acoustic tweezers applications, it is often desirable to trap a single particle and not its neighbors. Such selective capture requires reducing the spatial extent of the trapping region. Here, this can be conveniently achieved by decreasing the pulse width parameter $\sigma$. We note that this time-dependent strategy comes on top of existing solutions such as acoustic vortices \cite{baresch2016observation,riaud2017selective,baudoin2019folding,gong2020three}, and the two methods can therefore reinforce each-other.

Although using increasingly narrow pulses reduces the spatial extent of the wave, it does not necessarily improve its selectivity. Indeed, the \arf changes sign depending on the particles contrast factor \cite{settnes2012forces,karlsen2015forces}. Cells and particles have a positive contrast factor, meaning that these particles are trapped at the local minima of $\frac{\avg{\mathcal{U}}}{\Psi\avg{\mathcal{E}_\infty}}$. However, the energy landscape shown in \reffig{fig: U} has multiple local minima, which may result in as many trapping positions. A better measure of selectivity $\Delta\zeta_{max}$ is then how strongly the particles are bond to their local trap, especially the difference of restoring force between the strongest trap (primary trap) and the second strongest one (secondary trap)\footnote{The ratio of \arf could also have been considered here, but yields misleading results when the restoring force of the primary and secondary traps becomes vanishingly small.}. The variation of selectivity depending of the pulse width $\sigma$ is shown in \reffig{fig: selectivity}. Within the limits of the theory (particle much smaller than the highest frequency in the spectrum, which ultimately increases with $\sqrt{\sigma}$), the trapping selectivity of negative $\Psi$ particles always increases for narrower traps. However, positive contrast particles have an optimum selectivity for $\sigma=0.033\omega^2$. 

In order to compare this value, obtained for a pair of Gaussian wave packets, to the value used by Collins \etal~in their experimental study \cite{collins2016acoustic} with a triangular wave, we use a least square fitting to find the best approximation for $\sigma$. For a triangular wave-packet with $n_T$ periods, we get $\frac{\sigma}{\omega^2} \simeq 10.04 \frac{1}{2\pi {n_T}^2}$. For the 10-periods wave-packet used in their experiments, we get  $\sigma \simeq 0.016\omega^2$. According to \reffig{fig: selectivity}, this suggests that reducing the pulse duration twofold would not only yield a more localized trap but would also double the difference of force between primary and secondary traps. This is a very strong incentive to develop acoustofluidic devices in the pulsed regime using broadband acoustic transducers.

\section{Conclusion}

Acoustic radiation force has long been considered a steady-state time-averaged phenomenon. Here, we derive an expression of the \arf acting on spheres for arbitrary acoustic pulses, as long as the pulse bandwidth satisfies (i) the small sphere condition $f_{max} \ll \frac{c_0}{a}$ and (ii) pulse durations are short enough to neglect the sphere displacement compared to the wavelength $f_{min} \simeq 1/\tau \gg f_{max}\epsilon^2$. We provide an extension of the Gor'kov formula in the case of wave packets and other unsteady acoustic fields. Even in the time-domain, traveling waves in an inviscid fluid do not generate a net force up to $a^6$, while standing waves do yield a force proportional to $a^3$. Based on the complex monopole and dipole scattering coefficients, our model can account for thermoviscous effects that can considerably enhance the \arf of traveling waves. This model provides a theoretical foundation for the use of pulsed acoustic waves to enhance acoustic tweezers selectivity, and clarifies that an \arf exists even for a single acoustic period.

\begin{acknowledgments}
	
This work was supported by the National Natural Science Foundation of China with Grant No. 51950410582, 61874033 and 61674043, the Science Foundation of Shanghai Municipal Government with Grant No. 18ZR1402600, the State Key Lab of ASIC and System, Fudan University with Grant No. 2018MS003 and 2020KF006, ISITE-ULNE (ERC Generator program) and Institut Universitaire de France.
	
\end{acknowledgments}



\bibliography{transient_arf}

\end{document}